\documentclass[preprint2]{aastex}

\slugcomment{vs. 6: 2011 Oct 17 - Teddy's cosmetics + referee report}

\shorttitle{The Site of TeV Flaring in M87}
\shortauthors{Harris et al.}

\begin{document}


\title{An Experiment to Locate the Site of TeV Flaring in M87}


\author{D. E. Harris\altaffilmark{1}, F. Massaro\altaffilmark{1,15},
C. C. Cheung\altaffilmark{2,3},
D. Horns\altaffilmark{4,5},
M. Raue\altaffilmark{4,5},
\L. Stawarz\altaffilmark{5,6},\\
S. Wagner\altaffilmark{5,7},
P. Colin\altaffilmark{8,9},
D. Mazin\altaffilmark{9,10},
R. Wagner\altaffilmark{8,9},
M. Beilicke\altaffilmark{11,12},\\
S. LeBohec\altaffilmark{12,13},
M. Hui\altaffilmark{12,13}, \and
R. Mukherjee\altaffilmark{12,14}}



\altaffiltext{1}{SAO, 60 Garden St., Cambridge, MA 02138}

\altaffiltext{2}{National Research Council Research Associate, National Academy of
Sciences, Washington, DC 20001, resident at Naval Research Laboratory,
Washington, DC 20375, USA}

\altaffiltext{3}{a member of the Fermi-LAT collaboration}

\altaffiltext{4}{University of Hamburg,
Institute for Experimental Physics,
Luruper Chaussee 149,
D-22761 Hamburg,
Germany}

\altaffiltext{5}{a member of the H.E.S.S. collaboration}

\altaffiltext{6}{Institute of Space and Astronautical Science, JAXA, 3-1-1
Yoshinodai, Sagamihara, Kanagawa, 229-8510, Japan and Astronomical
Observatory, Jagellonian University, ul. Orla 171, 30-244 Krak\'ow,
Poland}

\altaffiltext{7}{Landessternwarte Heidelberg-Koenigstuhl,
69117 Heidelberg Germany}

\altaffiltext{8}{Max-Planck-Institut f\"ur Physik, D-80805 M\"unchen, Germany}

\altaffiltext{9}{a member of the MAGIC collaboration}

\altaffiltext{10}{IFAE, Edifici Cn., Campus UAB, E-08193 Bellaterra, Spain}    

\altaffiltext{11}{Washington University in St.Louis, Campus Box 1105, 1 Brookings
Drive, St.Louis, MO 63130}

\altaffiltext{12}{a member of the VERITAS collaboration}                       

\altaffiltext{13}{University of Utah, Department of Physics
115 South 1400 East,
Salt-Lake-City, UT 84112-0830} 

\altaffiltext{14}{Barnard College, Physics \& Astronomy, New York NY 10027}          

\altaffiltext{15}{SLAC National Laboratory and Kavli
Institute for Particle Astrophysics and Cosmology, 2575 Sand Hill
Road, Menlo Park, CA 94025}    


\begin{abstract} 
We describe a Chandra X-ray Target of Opportunity project designed to
isolate the site of TeV flaring in the radio galaxy M87.  To date, we
have triggered the Chandra observations only once (2010 April) and by
the time of the first of our 9 observations, the TeV flare had ended.
However, we found that the X-ray intensity of the unresolved nucleus
was at an elevated level for our first observation.  Of the more than
60 Chandra observations we have made of the M87 jet covering 9 years,
the nucleus was measured at a comparably high level only 3 times.  Two
of these occasions can be associated with TeV flaring, and at the time
of the third event, there were no TeV monitoring activities.  From the
rapidity of the intensity drop of the nucleus, we infer that the size
of the emitting region is of order a few light days $\times$ the
unknown beaming factor; comparable to the same sort of estimate for
the TeV emitting region.  We also find evidence of spectral evolution
in the X-ray band which seems consistent with radiative losses
affecting the non-thermal population of the emitting electrons within
the unresolved nucleus.
\end{abstract}



\section{Introduction: The Problem}\label{sec:intro}

Although the radio galaxy M87 is normally a weak source at TeV energies, occasionally
there is flaring activity which lasts from a few days up to a week or two.
During these times, the observed intensity can peak at $\geq$ 10\%
Crab.  To date there have been 3 well documented flarings:

\begin{itemize}

\item{2005 April - H.E.S.S. found variable TeV emission with typical
  timescales of a few days \citep{ahar06}.  Since this event coincided
  with the peak of the giant flare (radio/UV/X-ray) from the knot
  \mbox{HST-1} which lies at 0.86$^{\prime\prime}$ (60pc in
  projection) from the nucleus of M87, there has been an on-going
  debate as to whether the TeV emission originated near the
  supermassive black hole in the nucleus or from \mbox{HST-1}
  (\citet{harr09}, and references therein).}

\item{2008 Feb - A second flaring event which lasted for almost two
  weeks was observed by H.E.S.S., MAGIC, and VERITAS.  By chance, a
  series of VLBA observations was underway by R. C. Walker at 43~GHz:
  the milliarcsec nucleus progressively brightened \citep{acci09}.
  One of our standard Chandra X-ray monitoring observations occurred a
  few days after the TeV flaring and we found that the nucleus was
  brighter than usual whereas HST-1 maintained a comparatively
  constant intensity.  See also \citet{acci10}.}

\item{2010 Apr - This event was observed by H.E.S.S., MAGIC, and
  VERITAS \citep{ong10}.  Once the reality of the flaring was
  established, a Chandra target-of-opportunity (ToO) was triggered and
  these results are the subject of this paper.  Although \mbox{HST-1}
  has declined from its peak intensity in 2005 to levels similar to
  those in 2000, the nuclear X-ray emission was at a somewhat higher
  level than usual.  However, no activity at 43~GHz was detected.  For
  more details, see \citet{raue11}}.

\end{itemize}

From the TeV variability timescale, it is deduced that the TeV
emitting region is of the order of a light day times $\delta$, the
unknown Doppler beaming factor \citep{raue11}. Although we have not
found such short timescale variability in the X-rays from HST--1, the
nucleus has a somewhat shorter timescale for variability than the 20
days of \mbox{HST-1} \citep{harr09}.

So the open question is, can we determine the site of the TeV flaring by
identifying common features in the lightcurves at TeV and at X-rays?

At a distance of 16 Mpc, the angular scale for M87 is 77pc/arcsec.
Spectral indices are defined by flux density, S$_{\nu}\propto\nu^{-\alpha}$.

\section{The Experiment}\label{exp}
The basic idea for our Chandra ToO program on
M87 is to trigger a series of observations when the TeV exhibited a
level of $\geq$7\% Crab.  With luck, we had hoped that the X-ray light
curve of either the nucleus or \mbox{HST-1} would divulge some feature
that would correspond to the TeV excursions.  The main problem for
this experiment is the brevity of the TeV flaring compared to the time
it takes for Chandra mission planning to alter the observing schedule and
upload a revised observing schedule.

Our strategy was to make an initial 5 ks Chandra observation as soon
as possible after receiving the TeV trigger, and then to follow the
first with 4 more 5ks observations spaced at intervals of between 1.5
and 3 days.  We then had 4 more observations, the first of which would
start at the beginning of the next dark of the moon fortnight, and
these would be spaced by 3$\pm$1 days.  The lunar constraint was
imposed since it would make little sense to discover discrete features
in the X-ray light curves around the full moon when TeV observations
are difficult or impossible.

On Friday, 2010 April 9 the trigger level was realized (see
\citet{ong10}) and our first observation was late in the day on Sunday
(April 11).  Unfortunately, the final elevated TeV level occurred on
Saturday night; by Sunday night when we obtained the first Chandra
observation, the TeV flaring was over.

\section{The Observations \& Data Reduction}\label{sec:obs}

In Table~\ref{tab:obs} we list the particulars of our Chandra
observations.  In addition to the 9 observations of this ToO in 2010
April and May, we include a re-analysis of an observation (labeled
'Zh' in previous papers) obtained on 2008 Feb 16.5, which was 3 days
after the final activity (Feb 13) of an earlier TeV flaring event (see
\S\ref{sec:intro}).

\begin{deluxetable}{lrlc} 
\tablecaption{Chandra Observations\label{tab:obs}}
\tablehead{
\colhead{Name} & \colhead{OBSID} & \colhead{Start Time
    (UT)} & \colhead{Live Time (sec)} 
} 

\startdata 
Zhed & 8577  & 2008-02-16T11:30:12 & 4659 \\ 
eda1 & 11512 & 2010-04-11T21:14:39 & 4699 \\ 
eda2 & 11513 & 2010-04-13T14:16:43 & 4703 \\ 
eda3 & 11514 & 2010-04-15T20:32:42 & 4529 \\ 
eda4 & 11515 & 2010-04-17T21:47:42 & 4699 \\ 
eda5 & 11516 & 2010-04-20T13:20:52 & 4707 \\ 
eda6 & 11517 & 2010-05-05T19:25:21 & 4703 \\ 
eda7 & 11518 & 2010-05-09T02:39:49 & 4402 \\ 
eda8 & 11519 & 2010-05-11T11:17:03 & 4706 \\ 
eda9 & 11520 & 2010-05-14T09:04:59 & 4595 \\ 
\enddata

\tablecomments{The first column provides a label for each observation.
  'Zhed' combines the previous designation (Zh) with 'ed' which is
  short for EDSER, the subpixel repositioning algorithm used for all
  our data in this paper. 'eda' is a compression of 'EDSER April'.}
\end{deluxetable}

\subsection{Data Processing}

Our data processing of M87 5 ks observations (we have accumulated over
60 of these during the years 2002-2009; \citep{harr09,acci10}) has
purposefully remained essentially unchanged except for the changes
made to the CALDB.  In brief, we check for high background levels,
remove pixel randomization (the intentional degradation of image
quality by adding a random number to the pixel location of each event;
this process was a part of the pipeline processing until quite
recently), register the X-ray event file by changing the values of
appropriate keywords in the fits header so as to align the X-ray
nucleus with the radio nucleus, and then construct fluxmaps
regridded to one tenth of the native ACIS pixel size (i.e. pixels size
changes from 0.492$^{\prime\prime}$ to 0.0492$^{\prime\prime}$).  This
last step is done for the soft band (0.2-0.75 keV), the medium band
(0.75-2 keV), and the hard band (2-6 keV).  Because of severe
pileup\footnote{The term 'pileup' is used for CCD detectors to
  describe the situation of a detected event consisting of two or more
  different photons which happened to arrive during the same frame
  integration time and close to the same location on the detector.}
for the knot HST-1 during its high intensity flaring in 2005, we
forsook cgs units (the fluxmaps) and instead made our primary
intensity measurement in terms of keV/s.  This photometry was
performed using the event 1 file\footnote{The event 1 file contains
  all the events; the event 2 file has been filtered for standard
  grades and also filtered for good time intervals.}  with no grade
filtering, and integrated from 0.2 keV up to whatever energy was
required to recover essentially all the events associated with a given
feature.  In the rest of this paper, it is important to distinguish
between counts per sec (used primarily in dealing with evaluation of
pileup) and keV/s, which is the sum of each event multiplied by its
energy.  Since most photons are of order 1 keV, the magnitudes of
these two observational parameters are quite similar.  Further details
of our processing methods can be found in Papers I, III, and V of our
series \citep[][respectively]{harr03,harr06,harr09}.

\begin{figure} 
\plotone{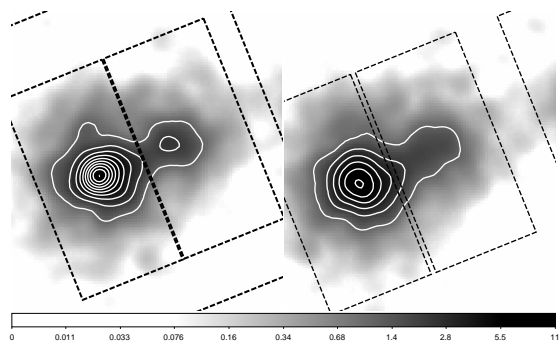}
\caption{A comparison of EDSER (left) vs. 'pixel randomization
  removed' (right) processing for the first observation of 2010 April
  ('eda1').  The nucleus and HST-1 (the feature towards the NW) are
  separated by 0.86$^{\prime\prime}$.  Each image has been rebinned to
  1/16 native ACIS pixel and smoothed with a Gaussian of FWHM =
  0.22$^{\prime\prime}$.  The grayscales are logarithmic but the
  contours are linear.  For the left panel (EDSER) the peak intensity
  is 10.648 counts/pixel and the contours are set to 10, 20, 30,
  ... 100\%.  The right panel has peak intensity of 6.59 counts/pixel
  and, for illustrative purposes, we have applied the same (absolute)
  contour levels, i.e. 1.07, 2.15, 3.23, ... 6.47 (cts/pix).  As
  described in the text. our standard rectangular photometric regions
  for the nucleus and HST-1 were found to be slightly misplaced when
  judged with an EDSER image.  In the right panel we show the standard
  rectangles and in the left panel are the slightly shifted
  rectangles.
\label{fig:eda1}}
\end{figure}

For the current effort we have altered our procedures in a few
particulars.  First we changed the 'bad pixel' routine to the 'hot
pix' recently released in CIAO.  More importantly however, we used
{\it acis\_process\_events} to apply the EDSER algorithm\footnote{The
  term 'EDSER' stands for ``energy dependent sub-pixel event
  repositioning'': it is an algorithm recently included in CIAO and
  the pipeline processing which uses the charge distribution within a
  3x3 pixel region to refine the assumed location of the incoming
  photon in detector coordinates.} in place of the removal of pixel
randomization.  In various tests, we found that data processed with
EDSER produce a more narrow point-spread function and consequently a
higher peak intensity.  While not a drastic improvement, it is
sufficiently good so as to be a valuable asset in separating the
nucleus from HST-1 (separated by 0.86$^{\prime\prime}$).  An example
(our first observation, 'eda1') is shown in fig.~\ref{fig:eda1}, which
will also serve to illustrate the apertures for photometry.

\subsection{Examination of Pileup}

Before attempting to evaluate the characteristics of the nuclear data
that exceed 1 kev/s, it behooves us to consider the effects of pileup
in the ACIS CCD's.

Although we have had to deal with severe pileup for the high intensity
of the knot HST-1 in 2005, we generally considered levels of $\leq$1
keV/s as weak enough so that pileup could be ignored.  However, in the
present case, we realize that even if a minority of incoming photons
arrived during a frame time with another photon, the resulting events
would masquerade as single photons of higher energy, distorting the
spectral distribution.  Therefore, to understand the effects of mild
pileup, we examined 3 parameters indicative of pileup, and use the
countrate of the event 1 file as our independent variable.

The first is the ratio of the measured intensities (keV/s) in the
event 2 file to that in the event 1 file.  We time filtered the event
1 file but did not filter for standard grades since we wanted to
recover those events which suffered grade migration,\footnote{The term
  'grade migration' describes what happens when a single event
  consists of more than one photon.  Many of these events (because the
  charge of the second or third photon is positioned at a different
  location) will cause the filtering software to label these as 'bad
  grades' (e.g. cosmic rays), and when grade filtering between event 1
  and event 2 files occurs, these events will be rejected.} a common
symptom of pileup.  For knot A which is slightly resolved and not
expected to suffer pileup, the intensity is about 0.2 c/s (0.08 counts
per frame since we always use 0.4s frame time).  Values for the ratio
keV/s(evt2/evt1) range between 0.95 and 0.98 for the 10 observations
of Table~\ref{tab:obs}.  For knot D which has an intensity roughly
half that of knot A, the observed ratios go from 0.96 to 0.99.

For the 9 observations of 2010, HST-1 had an intensity between 0.25
and 0.28 c/s and the observed ratio ranged from 0.89 to 0.95.  In
2008 Feb the intensity was 0.82 c/s and the ratio was 0.79.  For the
nucleus, the two intensities greater than 1 keV/s had ratios of 0.78
and 0.85 whereas the lower intensity observations ranged from 0.91 to
0.93 for the ratio.

\begin{figure}
\includegraphics[angle=-90,scale=.4]{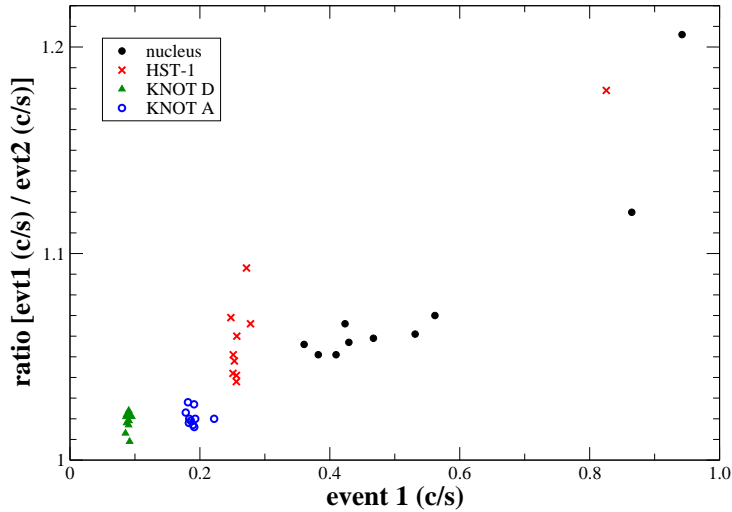}
\caption{The ratio of event 1 countrate to that from the event 2 file.
  When pileup becomes significant, the event 2 countrate decreases
  relative to the event 1 countrate as more and more events are
  rejected by the standard grade filtering.  The data point for the
  nucleus in 2008 Feb. has the largest ratio and the largest event 1
  intensity.  The other nuclear point at the right in the plot is from
  the eda1 observation, our first after the TeV flaring of 2010 April.
  HST-1 had a larger intensity (and hence a larger ratio) in 2008 than
  in 2010.  The data for knots D and A have ratios close to unity and
  are essentially free of pileup effects.  The fact that not all the
  data for the nucleus and HST-1 lie on a single curve is most likely
  caused by an interaction between these two components (e.g. a mild
  case of 'Eat Thy Neighbour', as described in
  \citet{harr09}).\label{fig:pileup}}
\end{figure}

The loss of detected energy passing from event 1 to event 2 is
reflected in the loss of detected events caused by grade migration.
In fig.~\ref{fig:pileup} we show this effect.

Another indicator of pileup is the average energy of all detected
events.  Although $<E>$ is ambiguous since it will increase if the
intrinsic source spectrum hardens, if coupled with detectable grade
migration, it is a necessary consequence of pileup.  Not only will the
energy of piled events be much larger than the average of unpiled
events, but the average energy will be augmented because there are
fewer events than there should be owing to the doubling up of piling.
This predicted effect is present in our data: the mean energies for
the nucleus for the 8 observations of intensity $<$ 1 keV/s range
between 1.38 and 1.53 keV while for Zhed it is 1.78 and for eda1 it is
1.57.  Similarly, $<E>$(Zhed) for HST-1 is 1.47 keV compared to the 2010
values which range between 1.15 and 1.28.

We have also used
PIMMS\footnote{http://heasarc.gsfc.nasa.gov/Tools/w3pimms.html} to
estimate pileup.  Specifying the input as Chandra c/s with frame time
0.4s, a power law with photon index = 2, and absorption by a column
density of $N_h=4\times10^{20}$cm$^{-2}$; we constructed a 'look-up'
table of input (photons per sec arriving at the detector) and output
c/s, which is the countrate of the unpiled photons.  We calculated the
total event rate from the expression:

\smallskip

observed(evt1 c/s) = $\frac{PIMMS(output) c/s}{1-Fp}$\\

\noindent
where Fp is the fractional pileup reported by PIMMS and is defined as
the number of piled events divided by the total number of events.  We
use the observed event rate from the event 1 file which has been time
filtered but not grade filtered because PIMMS does not accommodate
corrections for grade migration.  In this way we can estimate the
inferred counting rate from each component in the absence of pileup.
Relevant numbers are given in Tables~\ref{tab:nucdata} (nucleus)
\&~\ref{tab:hst1data} (HST-1), and it is clear that even with a frame time of
0.4s, pileup will significantly affect most source parameters except
for the intensity in keV/s, which was specifically designed to
circumvent first order pileup problems.

\begin{deluxetable}{lrrrrrl} 

\tablecaption{X-ray intensities, pileup
  parameters, and spectral indices for the
  nucleus.\label{tab:nucdata}}

\tablewidth{0pt} 

\tablehead{ 
  \colhead{Observation} & \colhead{Intensity}\tablenotemark{a} &
  \colhead{Event2} & \colhead{Event1} & \colhead{PIMMS}\tablenotemark{b} &
  \colhead{Fractional}\tablenotemark{c} & \colhead{$\alpha_x$}\\ 

 & \colhead{(keV/s)} & \colhead{(c/s)} & \colhead{(c/s)} &
  \colhead{(c/s)} & \colhead{Pileup} & \\ 
}

\startdata 

Zhed & 1.692$\pm$0.026 & 0.781 & 0.942 & 1.11 & 0.16 & 0.87$\pm$0.05 \\
eda1 & 1.365$\pm$0.021 & 0.772 & 0.865 & 1.01 & 0.15 & 0.92$\pm$0.04 \\
eda2 & 0.774$\pm$0.015 & 0.525 & 0.562 & 0.62 & 0.10 & 1.19$\pm$0.06 \\
eda3 & 0.641$\pm$0.015 & 0.406 & 0.429 & 0.46 & 0.07 & 1.02$\pm$0.07 \\
eda4 & 0.669$\pm$0.014 & 0.441 & 0.467 & 0.50 & 0.08 & 1.13$\pm$0.07 \\
eda5 & 0.650$\pm$0.015 & 0.397 & 0.423 & 0.45 & 0.07 & 1.06$\pm$0.07\\
eda6 & 0.767$\pm$0.015 & 0.501 & 0.531 & 0.58 & 0.09 & 1.05$\pm$0.05\\
eda7 & 0.577$\pm$0.014 & 0.390 & 0.410 & 0.45 & 0.07 & 1.12$\pm$0.07\\
eda8 & 0.541$\pm$0.013 & 0.364 & 0.382 & 0.41 & 0.06 & 1.06$\pm$0.07\\
eda9 & 0.517$\pm$0.013 & 0.341 & 0.360 & 0.38 & 0.06 & 1.03$\pm$0.08\\

\enddata

\tablecomments{The statistical uncertainties for the countrates are of
  order a few percent}

\tablenotetext{a}{The $\pm$ values are the statistical uncertainties
  derived from the corresponding countrates.}

\tablenotetext{b}{The PIMMS countrate attempts to estimate the
  rate at which photons are arriving at the detector, i.e. source
  count rate in the absence of pileup.}

\tablenotetext{c}{The fractional pileup from PIMMS, defined as the
ratio of piled events to total events.}

\end{deluxetable}

\begin{deluxetable}{lrrrrrl} 
\tablecaption{X-ray intensities, pileup parameters, and spectral indices for
  HST-1.\label{tab:hst1data}} 

\tablewidth{0pt}

\tablehead{ 
  \colhead{Observation} & \colhead{Intensity}\tablenotemark{a} &
  \colhead{Event2} & \colhead{Event1} & \colhead{PIMMS}\tablenotemark{b} &
  \colhead{Fractional}\tablenotemark{c} & \colhead{$\alpha_x$}\\ 

 & \colhead{(keV/s)} & \colhead{(c/s)} & \colhead{(c/s)} &
  \colhead{(c/s)} & \colhead{Pileup} & \\ 
}

\startdata 

Zhed & 1.212$\pm$0.020 & 0.700 & 0.826 & 0.95 & 0.14 & 1.16$\pm$0.05\\
eda1 & 0.348$\pm$0.010 & 0.249 & 0.272 & 0.28 & 0.04 & 1.32$\pm$0.10 \\
eda2 & 0.309$\pm$0.009 & 0.232 & 0.248 & 0.25 & 0.04 & 1.51$\pm$0.12 \\
eda3 & 0.292$\pm$0.009 & 0.242 & 0.253 & 0.26 & 0.04 & 1.53$\pm$0.12\\
eda4 & 0.316$\pm$0.009 & 0.246 & 0.256 & 0.26 & 0.04 & 1.67$\pm$0.12\\
eda5 & 0.343$\pm$0.009 & 0.261 & 0.278 & 0.29 & 0.04 & 1.49$\pm$0.11\\
eda6 & 0.308$\pm$0.009 & 0.242 & 0.257 & 0.26 & 0.04 & 1.56$\pm$0.12\\
eda7 & 0.306$\pm$0.009 & 0.239 & 0.251 & 0.26 & 0.04 & 1.52$\pm$0.13\\
eda8 & 0.312$\pm$0.009 & 0.241 & 0.251 & 0.26 & 0.04 & 1.56$\pm$0.11\\
eda9 & 0.301$\pm$0.009 & 0.247 & 0.256 & 0.26 & 0.04 & 1.56$\pm$0.11 \\

\enddata


\tablecomments{The statistical uncertainties for the countrates are of
  order a few percent}

\tablenotetext{a}{The $\pm$ values are the statistical uncertainties
  derived from the corresponding countrates.}

\tablenotetext{b}{The PIMMS countrate attempts to estimate the
  rate at which photons are arriving at the detector, i.e. source
  count rate in the absence of pileup.}

\tablenotetext{c}{The fractional pileup from PIMMS, defined as the
ratio of piled events to total events.}

\end{deluxetable}

\subsection{Photometry}

\begin{figure}   
\epsscale{1.1}
\plotone{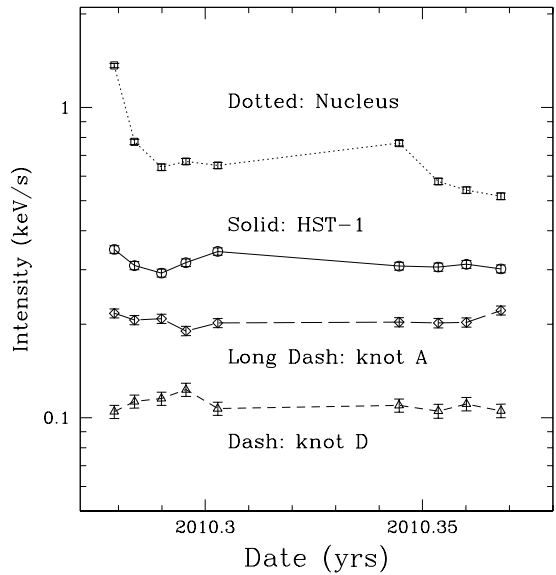}
\caption{The light curves for (top to bottom) the nucleus, HST-1, knot
  A, and knot D.  Intensity units are keV/s and are derived from the
  event 1 file with no grade filtering.  The first observation was
  obtained on 2011 April 11; the last on May 14.\label{fig:lc}}
\end{figure}

X-ray intensities were measured using the rectangular apertures defined
previously.  The keV/s values are given in Tables~\ref{tab:nucdata}
\&~\ref{tab:hst1data}, and plotted in fig.~\ref{fig:lc}.  We do not
include tabular data for knots D and A since they are not viable
candidates for the site of flaring TeV emission and pileup is negligible for
both: knot D has a much smaller counting rate than the nucleus and
HST-1, whereas knot A is resolved.  The data for these two knots are
included in fig.~\ref{fig:lc} as 'control features'; any intrinsic
variability in these two knots is at a much lower level than for the
nucleus and HST-1.

\section{Discussion}

If the predominant emission mechanism for the nuclear X-rays is
synchrotron emission as it is for \mbox{HST-1}, then there must be a
sizable population of high energy relativistic electrons whose IC
emission will be in the TeV range \citep{harr09}.  In
fig.~\ref{fig:lc} we show the light curves for the nucleus,
\mbox{HST-1}, knots D and A. Note the high level of the nucleus at our
first observation and the sharp drop to more normal levels by the time
of the second observation.

Although our experiment failed to isolate discernable features in the
X-ray light curves corresponding to TeV features, it yielded
tantalizing hints that we most likely witnessed the tail end of a
period of X-ray flaring associated with the TeV flaring.  The
intriguing question is what we might have seen if Chandra observations
had begun a few days earlier.  If the slope of the nuclear light curve
before our first observation was similar to that between the first two
observations, the X-ray intensity of the nucleus during the TeV
flaring would have been remarkable.

During our 8 years of monitoring the M87 jet with Chandra
\citep{harr09,acci10}, the nuclear emission has seldom been larger
than 1 keV/s: on two occasions (2008 Feb 16 and 2010 Apr 11) these
levels were in close proximity to TeV flaring.  A third time (2006 Jun
28), there was no TeV coverage, and for the 2005 April TeV flaring,
our data are not reliable because of second order pileup effects
associated with \mbox{HST-1}.

\subsection{The X-ray - TeV Connection}

For the case of HST-1, we became convinced that the huge flare of 2005
could be described by a broadband synchrotron emission from the radio
to the X-rays in an equipartition field of $\approx$1mG (Papers I,
III, \& V).  With the presence of electrons with Lorentz factors,
$\gamma\approx~10^7$, we showed that such a source would be expected
to produce inverse Compton (IC) emission in the TeV range from ambient
photons (Paper IV).  Moreover, although it is probable that expansion
losses were operative, there were time intervals which manifest the
signature of energy-dependent radiative losses and the decay times
were consistent with a 1mG field (Paper V).

When we consider the emission from the unresolved (by Chandra) nuclear
component, we cannot devise a complete scenario because there are so
many unknowns: primarily the source size and the broadband nature of
the emission.  It may be the case that the operative field strength is
significantly larger than that of HST-1, the emitting volume may be
much smaller, and the beaming factor could be quite different (either
larger or smaller than the estimate of $\delta\approx$~5 for HST-1;
\citet{perl11}).  In \citet{raue11} it is clearly shown that the
evidence at hand provides mixed results when attempting to show
broadband emission associated with TeV flaring.  While one can argue
that the X-ray flux observed from the nucleus is always a bit larger
than normal at the time of TeV flaring, the 43 GHz 'activity' from the
milliarcsec core appeared on one occasion but was absent on another.

One might, in fact, propose that the major acceleration mechanism could
be reconnection rather than shock acceleration, and therefore, there
could well be quite a different distribution of relativistic electrons
from the broken power law from low values of $\gamma$ up to 10$^7$ or
10$^8$ suggested to explain the broadband emission from HST-1.  If
such were to be the case, there is no a priori reason to expect that a
particular band will show anomalous emission directly associated with
the TeV flaring.  With this caveat in mind, we will now argue that
there is, in fact, circumstantial evidence for X-ray emission from the
nucleus of M87 which is associated with TeV flaring.

\subsection{The 2010 April X-ray Event}\label{sec:april}

For the nucleus, the spectral differences we have measured between the
first and remaining observations are in the same direction as would be
expected from the change in pileup fraction.

The drop in the nuclear intensity between our first two observations
(separated by 1.68 days) permits us to measure the fractional change
per day, '{\it fpd}' (see Paper V where {\it fpy} - 'fractional change per year' was
introduced) and compare its value to the historical behavior during 8
years of monitoring.  Typically, the largest values found previously
for the nucleus were {\it fpy} $\approx$50, corresponding to an {\it fpd} value of 
{\it fpy}/365 $\approx$0.137 (both positive and negative {\it fpy's} of
this magnitude were observed).  The {\it fpd} (or {\it fpy}) formulae are:

\smallskip

$fpd=(I_2-I_1)/(I_i\Delta~t)$\\

$\sigma(fpd) = \frac{1}{\Delta(t)}\times\frac{I_j}{I_i}\times\sqrt{(\frac{\sigma_1}{I_1})^2~+~(\frac{\sigma_2}{I_2})^2}$

\noindent 
where $I_1$ and $I_2$ are successive intensities,
$\Delta~t=(t_2-t_1$), and i=1, j=2 when the source is getting
brighter and i=2, j=1 when the intensity is dropping.
The time it would take to change the intensity by a factor of two
for an observed value of {\it fpd} is 1/{\it fpd}.

The calculated values of {\it fpd} (between eda1 and eda2) for various
intensity measurements are $-$0.454$\pm$0.027 for keV/s;
$-$0.150$\pm$0.041 for the soft flux; $-$0.266$\pm$0.030 for the medium
flux; and $-$0.643$\pm$0.071 for the hard flux.  While it would be
useful to know if the hard flux was dropping faster than the soft
flux, the apparent effect could be caused, in whole or in part, by the
reduction of the pileup fraction between eda1 and eda2.

To estimate the spectral distribution of the incident photons
(i.e. what we would have observed in the absence of pileup), we
adopted the rather extreme assumption that all the piled events ended
up in the hard band, and assumed that of the arriving photons which
eventually piled, 1/3 came from the soft band and 2/3 came from the
medium band.  This is 'extreme' because some of the piled events
will actually end up in the medium band and because some fraction of
the piled events will actually consist of 3 photons rather than just
two.  Taking the observed count rates from the event 1 file (i.e. no
grade filtering) we were then able to estimate the {\it incident}
count rates for each band for both eda1 and eda2.  It is then a simple
matter to calculate {\it fpd} values but the uncertainties will be dominated
by errors in assuming the particulars of the pileup.  Therefore, we
have calculated the uncertainties for the observed {\it fpd}, but not for
the assumed incident count rates: the results are presented in
Table~\ref{tab:fpy}.

\begin{deluxetable}{lrrrr} 
\tablecaption{The X-ray Intensity Decay of the Nucleus \label{tab:fpy}}

\tablehead
{
& \colhead{Soft} & \colhead{Medium} & \colhead{Hard} & \colhead{Total} \\
& \colhead{0.2-0.75keV} & \colhead{0.75-2keV} & \colhead{2-6keV} & \colhead{0.2-6keV} \\
}

\startdata

eda1(observed) c/s &  0.167 &  0.498 &  0.192 &  0.857 \\
eda2(observed) c/s &  0.139 &  0.335 &  0.085 &  0.559 \\
{\it fpd}(observed)\tablenotemark{a}      &  $-$0.122 & $-$0.288  & $-$0.750  & $-$0.317\\
$\sigma$(fpd-obs)\tablenotemark{b}  &  0.038  &  0.029  &  0.081  &  0.023\\
eda1(inferred) c/s &  0.257 &  0.678 &  0.055 &  0.991 \\
eda2(inferred) c/s &  0.179 &  0.405 &  0.028 &  0.614 \\
{\it fpd}(inferred)\tablenotemark{a}  & $-$0.262  & $-$0.400  & $-$0.587  & $-$0.366\\

\enddata

\tablecomments{The observed values come from the event 1 file (no
  grade filtering).  The 'inferred' entries are estimated incident
  count rates (i.e. in the absence of pileup).}

\tablenotetext{a}{{\it fpd} is the fractional change per day.  The values
  quoted in \S\ref{sec:april} are based on flux maps; here in the
  table they are based on count rate.}

\tablenotetext{b}{$\sigma$(fpd-obs) is an estimate of the uncertainty
  of {\it fpd}, calculated in the standard fashion ($\sqrt{N}$/N, where N is
  the number of events in the appropriate energy band).}

\end{deluxetable}

The 'exercise' of assuming how the piled events are distributed
among the energy bands is just to get some rough idea to
differentiate between pileup effects and intrinsic attributes. 
{\it  Fpd} is a good measure of the rate of decay of intensity and we take
the decrease of {\it fpd}(incident) moving from soft to hard as
evidence for the presence of energy-dependent radiative losses.  The
fact that the spectral index for eda2 is the largest among all our
observations even though the intensity is second only to eda1, is
further evidence that pileup is not the dominant factor for the 
{\it fpd} dependency on energy.

Generally speaking, the most plausible mechanisms for a drop of
intensity for a non-thermal source is either expansion (which has no
first order effect on the spectral shape of the electron energy
distribution) or the radiative cooling with the energy loss rate
increasing with increasing energies of the radiating particles (for
example as $\propto E^2$ in the case of the synchrotron and inverse
Compton emission).
\footnote{Caveat: simple expansion can produce a similar spectral
  signature if there is a break in the power law spectrum or the
  spectrum is curved \citep{harr09}.}  Although the dimming of radio
emission from parsec scale knots is often ascribed to expansion, when
dealing with high energy emissions which must come from electrons with
much greater energies than those producing radio emission,
energy-dependent radiative losses are much more likely to
contribute to the dimming, with or without expansion.

From the general magnitude of {\it fpd} values we may also infer that the
light travel time across the X-ray emitting region is limited to a few
days, thus being consistent with estimates of the size of the TeV flaring
regions deduced by the timescale of TeV variability \citep{ahar06,raue11}.

Although our experiment did not succeed in isolating similar features
in the X-ray and TeV light curves, we find evidence that supports the
notion that there was concomitant flaring in the X-rays, and we
witnessed the last phase of that flaring.



\acknowledgments 

K. Mukai provided helpful information and advice on the implementation
of pileup in PIMMS.  We thank R. C. Walker for helpful suggestions on
the manuscript.  F. Massaro acknowledges the Fondazione Angelo Della
Riccia for the grant awarded him to support his research at SAO during
2011 and the Foundation BLANCEFLOR Boncompagni-Ludovisi, n'ee Bildt
for the grant awarded him in 2010 to support his research.  We thank
the {\it Chandra} mission planning team and the director's office for
their tireless efforts to reschedule target of opportunity
observations.  The anonymous referee is thanked for his useful
comments.  The work at SAO was supported by NASA grant GO0-11120X.



{\it Facilities:} \facility{Chandra}.





\end{document}